\documentclass[aps,pra]{revtex4}
\usepackage{graphicx}
\usepackage{tabularx}
\usepackage{multirow}
\usepackage{amsmath}

\begin{document}

\title{Single-qubit operation sharing with Bell and W product states}
\author{Qibin Ji$^a$, Yimin Liu$^b$, Xiansong Liu$^a$, Xiaofeng Yin$^a$, Zhanjun Zhang$^{a,\dag}$ \\
{\normalsize $^a$ School of Physics \& Material Science, Anhui
University, Hefei 230039, Anhui, China} \\
{\normalsize $^b$ Department of Physics, Shaoguan University, Shaoguan 512005, China} \\
{\normalsize $^{\dag}$zjzhang@ahu.edu.cn } }
\maketitle

\begin{minipage}{420pt}
{\bf Abstract}\ Two tripartite schemes are put forward with shared entanglements and LOCC for
sharing an operation on a remote target sate. The first scheme uses a Bell and a symmetric
W states as quantum channels, while the second replaces the symmetric W state by an asymmetric one.
Both schemes are treated and compared from the aspects of quantum resource consumption, operation
complexity, classical resource consumption, success probability and efficiency. It is found that,
the latter scheme is better than the former one. Particularly, the sharing can be achieved only
probabilistically with the first scheme deterministically with the second one.
\vskip 0.3cm
\noindent {\bf Keywords}: single-qubit operation sharing, Bell state, symmetric W state, asymmetric W state
\vskip 0.2cm
\noindent {\bf PACS numbers}: {03.65.Ta, 03.67.-a}
\end{minipage}\\\\

\noindent {\bf 1 \ Introduction}

Quantum entanglement is an important resource in various fields of quantum information processing,
such as quantum key distribution(QKD)[1-3], quantum state teleportation(QST)[4-7],
quantum secret sharing(QSS)[8-12] and quantum operation teleportation(QOT)[13-16], etc.
In 2011 Zhang and Cheung[17] have definitely put forward quantum operation sharing(QOS)
with the aid of local operation and classical communication (LOCC) as well as shared entanglements.
The basic idea of QOS in a simplest case is that, by virtue of shared entanglements and LOCC the
performer of a single-qubit operation can assure the operation be securely performed on a target
state in a remote agent's qubit if and only if both agents cooperate. Utilizing different
entangled states as quantum channels, such as Bell and GHZ states[17], five-qubit cluster state[18],
five-qubit Brown state[19], generalized Bell and GHZ qutrit states[20], etc[21-22], recently a variety
of QOS schemes have been proposed and this topic has attracted some attention.

W states were first presented by W. D$\ddot{u}$r et al[23] and have been extensively studied in the last decade.
As multi-particle entangled states, they have been exploited to fulfill various quantum tasks in different
quantum scenarios. Nowadays, it is well admitted that W states are a kind of important quantum resource
in quantum information processing. Due to some of their inherent advantages (e.g., their robustness),
W states have been attracting much attention[24-27] today. In this paper, we will use Bell states and two
different W states as shared entanglements to study the issue of QOS. Specifically, we will put forward
two different QOS schemes. One uses a Bell and a symmetric tripartite W states as quantum channels,
while the other utilizes a Bell and an asymmetric W states. After we present our proposals,
we will reveal both schemes' differences with respect to quantum resource consumption,
operation complexity, classical resource consumption, success probability and efficiency.

The rest of this paper is organized as follows. In section 2, we will present the two QOS
schemes, respectively. In section 3, we will show their important features, including security,
symmetry, probability and compare our two schemes in the five aspects: the quantum resource consumption,
the difficulty or intensity of necessary operations, the classical resource consumption,
the success probability and the intrinsic efficiency of the schemes. At last we will give a concise
summary in section 4. \\

\noindent {\bf 2 \ Two schemes with Bell states and two different W states}

In both two schemes there are three legitimate users, say, Grey, Holly and Jack. Let Grey
be the initial performer of the concerned operation $\Omega$, Holly and Jack be the two sharers.
Grey wants to perform a unitary operation $\Omega$ on the target qubit in one agent's site.
Actually, he may not hear of the concerned
operation $\Omega$ before, either. He wants to fulfill the task with his agents' assistance
and by making use of the quantum and classical channels linking he and agents. However, he does not
trust either agent completely. Specifically, he should certain that the operation can not be
successfully executed on the qubit by either agent solely but conclusively achieved via the mutual
collaboration of his two agents. Suppose the sharer Holly has the target qubit to be finally operated.
The qubit is in an arbitrary state and labeled as $h'$ reads
\begin{eqnarray}
|\Psi\rangle_{h'}= a|0\rangle_{h'} + b|1\rangle_{h'},
\end{eqnarray}
where $a$ and $b$ are arbitrarily complex and satisfy $|a|^2+|b|^2=1$.
To start with, we assume that Grey shares a Bell state
\begin{eqnarray}
|\psi^+\rangle_{g_1h_1}=\frac{1}{\sqrt{2}}(|00\rangle_{g_1h_1}+|11\rangle_{g_1h_1})
\end{eqnarray}
with Holly, where the qubit $g_1$ is in Grey's site, the qubit $h_1$ in Holly's position.

\vskip 0.2cm
\noindent {\bf 2.1 \ Scheme with Bell state and symmetric W state}

Now let us present our first QOS scheme (called as the S1 scheme later). The schematic demonstration is illustrated in figure 1. The
scheme can be concisely depicted as follows.
\vskip -1.0cm
\begin{figure}[h]
\begin{center}
\includegraphics[width=5.8in]{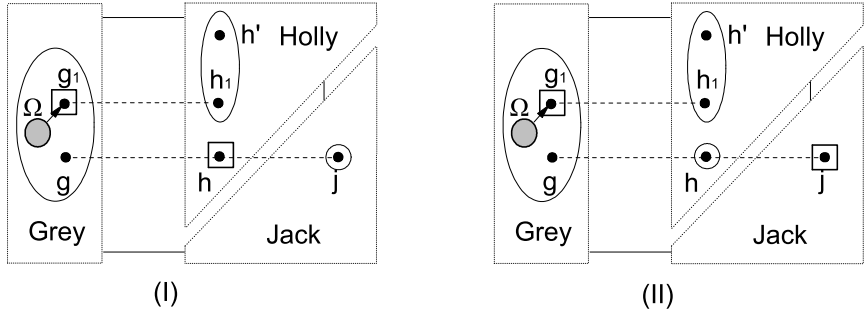}
\vskip -5.5cm
\begin{minipage}{350pt}
\caption{ Illustration of the three-party QOS scheme with a Bell state and a symmetric W state.
Dotted rectangle and trapeziums are participants' locations. Solid lines among rectangles stand
for classical channels. Solid dots denote qubits. Dash lines linking qubits are quantum channels.
Solid ellipses represent Bell-state measurements. Circle illustrates the single-qubit measurement.
Solid square illustrates unitary operation and Gray solid circle labels the unitary
operation $\Omega$, respectively. In (I) Holly can reconstruct the state $\Omega|\Psi\rangle$,
while in (II) Jack can. See text for more details.}
\end{minipage}
\end{center}
\end{figure}

Except for the Bell state, another shared entanglement employed by the three participants
is a symmetric W state
\begin{eqnarray}
|W_s\rangle_{ghj}=\frac{1}{\sqrt{3}}(|001\rangle_{ghj}+|010\rangle_{ghj}+|100\rangle_{ghj}),
\end{eqnarray}
where the qubits $g$, $h$ and $j$ are in the location of Grey, Holly and Jack, respectively.

At first Holly carries out a Bell-state measurement on his qubit pair $(h',h_1)$ and announces publicly the
outcome (two classical bits) via classical channel. Throughout this paper the four Bell states are
written as
\begin{eqnarray}
|\psi^{\pm}\rangle=(|00\rangle\pm|11\rangle)/\sqrt{2}, \ \ \
|\varphi^{\pm}\rangle=(|01\rangle\pm|10\rangle)/\sqrt{2}.
\end{eqnarray}
Obviously, the composite system consists of the channel qubits $(g_1, h_1)$ and the
initial target qubit $h'$, its state is
\begin{eqnarray}
|\Xi\rangle_{h'g_1h_1}=|\Psi\rangle_{h'}|\psi^+\rangle_{g_1h_1}=\frac{1}{\sqrt{2}}(a|0\rangle_{h'} + b|1\rangle_{h'})(|00\rangle_{g_1h_1}+|11\rangle_{g_1h_1}).
\end{eqnarray}
Naturally, Holly's measurements lead to the following collapses:
\begin{eqnarray}
|\psi^{+}\rangle_{h'h_1} \Rightarrow \sigma_{g_1}^x|\Psi\rangle_{g_1},\
|\psi^{-}\rangle_{h'h_1} \Rightarrow \sigma_{g_1}^y|\Psi\rangle_{g_1},\
|\varphi^{+}\rangle_{h'h_1} \Rightarrow |\Psi\rangle_{g_1},\
|\varphi^{-}\rangle_{h'h_1} \Rightarrow \sigma_{g_1}^z|\Psi\rangle_{g_1},
\end{eqnarray}
where $\sigma^x=|0\rangle\langle1|+|1\rangle\langle0|$ , $\sigma^y=|0\rangle\langle1|-|1\rangle\langle0|$
and $\sigma^z=|0\rangle\langle0|-|1\rangle\langle1|$ are Pauli operators.

Once receiving Holly's information, then Grey executes an appropriate Pauli operation
to convert the state in qubit $g_1$ to $|\Psi\rangle$ (see Table 1). To be specific, if Holly measures
$|\psi^+\rangle_{b'b_1}$, $|\psi^-\rangle_{b'b_1}$, $|\varphi^+\rangle_{b'b_1}$,
or $|\varphi^-\rangle_{b'b_1}$, Grey decides to perform $\sigma^x$, $\sigma^y$, $I$
or $\sigma^z$ on her qubit $g_1$, where $I$ is an identity operator.\\

\begin{center}
\begin{minipage}{360pt}
Table 1. Summary of the first stage in the first scheme.
HMO: Holly's measurement outcome.   HM: Holly's two-classical-bit message.
GCS: the collapsed state of qubit $g_1$ after Holly's measurement.
GO: the Pauli operation of Grey.   GRQS: the final recovered state of qubits $g_1$.
See text for more details. \\
\end{minipage}
\begin{tabular*}{13cm}{@{\extracolsep{\fill}}ccccc}
\hline
HMO                          & HM  & GCS                                 & GO                &GRQS                    \\ \hline
$|\psi^+\rangle_{h'h_1}$     & 00  & $\sigma_{g_1}^x|\Psi\rangle_{g_1}$  & $\sigma^x_{g_1}$  & $|\Psi\rangle_{g_1}$ \\
$|\psi^-\rangle_{h'h_1}$     & 01  & $\sigma_{g_1}^y|\Psi\rangle_{g_1}$  & $\sigma^y_{g_1}$  & $|\Psi\rangle_{g_1}$ \\
$|\varphi^+\rangle_{h'h_1}$  & 10  & $|\Psi\rangle_{g_1}$                & $I$               & $|\Psi\rangle_{g_1}$ \\
$|\varphi^-\rangle_{h'h_1}$  & 11  & $\sigma_{g_1}^z|\Psi\rangle_{g_1}$  & $\sigma^z_{g_1}$  & $|\Psi\rangle_{g_1}$ \\
\hline
\end{tabular*}
\end{center}

\vskip 0.6cm
Subsequently, Grey carries out the operation $\Omega$ on her qubit $g_1$,  i.e.,
\begin{eqnarray}
(\Omega|\Psi\rangle)_{g_1} = a'|0\rangle_{g_1} + b'|1\rangle_{g_1}. \ \
\end{eqnarray}
This indicates the concerned operation $\Omega$ has been performed on the target state $|\Psi\rangle$.

Next, Grey carries out a Bell-state measurement on his qubit pair $(g_1,g)$ and announces
publicly the outcome (two classical bits).
Note that, before Grey's measurement the total joint state of the four qubits $(g_1, g, h, j)$ is
\begin{eqnarray}
|\Gamma\rangle_{g_{1}ghj}&=&(\Omega|\Psi\rangle_{g_{1}})|W_s\rangle_{ghj} \\ \nonumber
&=&\frac{1}{\sqrt{3}}(a'|0\rangle_{g_1} + b'|1\rangle_{g_1})(|100\rangle_{ghj}+|010\rangle_{ghj}+|001\rangle_{ghj}).
\end{eqnarray}
It can be rewritten as
\begin{eqnarray}
|\Gamma\rangle_{g_{1}ghj}&=&\frac{1}{\sqrt{6}}[|\psi^{+}\rangle_{g_1g}(a'|01\rangle+a'|10\rangle+b'|00\rangle)_{hj}
+|\psi^{-}\rangle_{g_1g}(a'|01\rangle+a'|10\rangle-b'|00\rangle)_{hj} \nonumber\\
&&+|\varphi^{+}\rangle_{g_1g}(b'|01\rangle+b'|10\rangle+a'|00\rangle)_{hj}
+|\varphi^{-}\rangle_{g_1g}(-b'|01\rangle-b'|10\rangle+a'|00\rangle)_{hj}].
\end{eqnarray}
From this reexpression, one is readily to see that Grey's measurements
lead to the following collapses:
\begin{eqnarray}
\left\{ \begin{array}{*{20}{l}}
|\psi^+\rangle_{g_1g} \Longrightarrow a'|01\rangle_{hj}+a'|10\rangle_{hj}+b'|00\rangle_{hj},  \\
|\psi^-\rangle_{g_1g} \Longrightarrow a'|01\rangle_{hj}+a'|10\rangle_{hj}-b'|00\rangle_{hj},  \\
|\varphi^+\rangle_{g_1g} \Longrightarrow b'|01\rangle_{hj}+b'|10\rangle_{hj}+a'|00\rangle_{hj}, \\
|\varphi^-\rangle_{g_1g} \Longrightarrow -b'|01\rangle_{hj}-b'|10\rangle_{hj}+a'|00\rangle_{hj}.
\end{array} \right.
\end{eqnarray}

If Holly and Jack collaborate and decide Holly to conclusively reconstruct the conceivable state, then
they can do as follows. First, Jack measures his qubit $j$ with the computational bases $\{|0\rangle, |1\rangle\}$.
If $|0\rangle$ is measured, then he tells Holly the massages through their classical communication.
Otherwise, he does nothing. In the latter case, the sharing of the unitary operation has already failed
at this stage. Easily one can rewrite the right hand of equation (10) as
\begin{eqnarray}
\left\{ \begin{array}{*{20}{l}}
a'|01\rangle_{hj}+a'|10\rangle_{hj}+b'|00\rangle_{hj} = \sigma^x_h(\Omega|\Psi\rangle)_h|0\rangle_j + \alpha'|0\rangle_h|1\rangle_j,  \\
a'|01\rangle_{hj}+a'|10\rangle_{hj}-b'|00\rangle_{hj} = \sigma^y_h(\Omega|\Psi\rangle)_h|0\rangle_j + \alpha'|0\rangle_h|1\rangle_j,  \\
b'|01\rangle_{hj}+b'|10\rangle_{hj}+a'|00\rangle_{hj} = (\Omega|\Psi\rangle)_h|0\rangle_j + \beta'|0\rangle_h|1\rangle_c, \\
-b'|01\rangle_{hj}-b'|10\rangle_{hj}+a'|00\rangle_{hj} = \sigma^z_h(\Omega|\Psi\rangle)_h|0\rangle_j - \beta'|0\rangle_h|1\rangle_j.
\end{array} \right.
\end{eqnarray}
Secondly, after the transformation, if the measurement result is $|0\rangle$, then Holly can apply one of the
unitary transformations $\{\sigma_x, \sigma_y, I, \sigma_z\}$ to convert the state with Jack's help
(i.e., Jack tells him the result via their classical communication). By virtue of this scheme,
the unknown arbitrary operation can be shared by the sharer entity. Nonetheless, such situation
only appears with a certain probability. Easily one can work out each occurrence probability.
For $|\psi^+\rangle$, it is
\begin{eqnarray}
P_1=|_{g_1g}\langle\psi^+|\Gamma\rangle_{g_1ghj}|^2 = [1+|a'|^2]/6.
\end{eqnarray}
Analogously, probabilities of other outcomes $\psi^-$, $\varphi^+$ and $\varphi^-$ are
$[1+|a'|^2]/6$, $[1+|b'|^2]/6$ and $[1+|b'|^2]/6$, respectively.
Furthermore, the total probability that Holly measures $|0\rangle$ from the collapsed state
is $1/[1+|a'|^2]$. Consequently, the total success probability of this scheme is $2/3$.
For the sake of easy knowing about the correspondence, we have summarized all cases in table 2.

\vskip 0.2cm
\begin{center}
\begin{minipage}{390pt}
Table 2.  Summary of the second stage in the first scheme.
GMO: Grey's measurement outcome.   GM: Grey's two-classical-bit message.
HJCS: the collapsed state of qubits $h$ and $j$ after Holly's measurement.
JM: Jack's measurement outcome.    JM: Jack's single-classical-bit message.
HO: the Pauli operation of Holly.   HRQS: the final recovered state of qubits $h$.
See text for more details.
\end{minipage}
\end{center}
\begin{center}
\begin{tabularx}{13.5cm}{Xcc|XXXX}
\hline
GMO  & GM  & HJCS  & JMO  & JM   & HO   & HRQS  \\
\hline
\multirow{2}{*}{$\psi^+\rangle_{g_1g}$}  & \multirow{2}{*}{00} &\multirow{2}{*}{$a'|01\rangle_{hj}+(a'|10\rangle+b'|00\rangle)_{hj}$}  &$|0\rangle_j$  & 0
& $\sigma^x_h$    &$\Omega|\Psi\rangle_h$   \\
\cline{4-7}
&     &   &$|1\rangle_j$    &     & /   &   \\
\hline
\multirow{2}{*}{$\psi^-\rangle_{g_1g}$}  &  \multirow{2}{*}{01}   &\multirow{2}{*}{$a'|01\rangle_{hj}+(a'|10\rangle-b'|00\rangle)_{hj}$}  &$|0\rangle_j$  & 0
& $\sigma^y_h$    &$\Omega|\Psi\rangle_h$   \\
\cline{4-7}
&     &   &$|1\rangle_j$    &     & /   &  \\
\hline
\multirow{2}{*}{$\varphi^+\rangle_{g_1g}$}    &  \multirow{2}{*}{10}   &\multirow{2}{*}{$(b'|10\rangle+a'|00\rangle)_{hj}+b'|01\rangle_{hj}$}   &$|0\rangle_j$  & 0
& $I$   &$\Omega|\Psi\rangle_h$  \\
\cline{4-7}
&     &   &$|1\rangle_j$    &     & /  &  \\
\hline
\multirow{2}{*}{$\varphi^-\rangle_{g_1g}$}    &  \multirow{2}{*}{11}   &\multirow{2}{*}{$(-b'|10\rangle+a'|00\rangle)_{hj}-b'|01\rangle_{hj}$}   &$|0\rangle_j$  & 0
& $\sigma^z_h$   &$\Omega|\Psi\rangle_h$   \\
\cline{4-7}
&     &   &$|1\rangle_j$    &     & /  &   \\
\hline
\end{tabularx}
\end{center}

\vskip 0.5cm
If the two agents collaborate and decide Jake to conclusively reconstruct the conceivable state,
then they can fulfill the sharing in the almost same way as that Holly is chosen. The only
difference [cf. figure 1 (I) and (II)] is that Holly and Jack should exchange their performances.
That is, Holly measures his qubit $h$ with the computational bases. If he measures $|0\rangle$,
then tells Jack the massages via their classical channel. Otherwise, he does nothing. When Jack
receives Holly's message, Jack applies one of the unitary transformations $(\sigma_x, \sigma_y, I, \sigma_z)$
to recover the conceivable state $(\Omega|\Psi\rangle)$.

\vskip 0.2cm

\noindent {\bf 2.2 \ Scheme with Bell state and asymmetric W state}

Now let us move to propose our second QOS scheme (referred to as the S2 scheme hereafter).
The schematic demonstration is illustrated in figure 2.
The scheme can be concisely depicted as follows.

\begin{figure}[h]
\begin{center}
\includegraphics[width=5.9in]{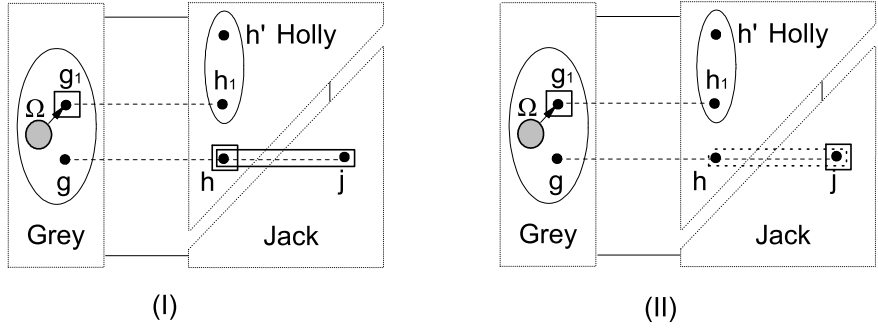}
\vskip -5.5cm
\begin{minipage}{370pt}
\caption{ Illustration of the three-party QOS scheme with a Bell state
and an asymmetric three-qubit W state. The same as figure 1 except that solid
and dotted rectangles represent sharers' different joint unitary operations,
respectively. See text for more details.}
\end{minipage}
\end{center}
\end{figure}

As mentioned before, in this scheme an asymmetric three-qubit W state is employed instead of the symmetric one
in the first scheme. Grey owns the qubit $g$, Holly the qubit $h$, and Jack the qubit $j$, respectively.
The asymmetric W state reads
\begin{eqnarray}
|W_a\rangle_{ghj} = \frac{1}{\sqrt{2}}(\alpha|001\rangle_{ghj} + \beta|010\rangle_{ghj} + |100\rangle_{ghj}),
\end{eqnarray}
where $\alpha$ and $\beta$ are complex and satisfy $|\alpha|^2 + |\beta|^2 = 1$.

As same as the first scheme, the Bell state is taken as quantum channel to realize the QST process,
and the concerned operation is implemented on the target state which has been teleported from the
qubit $h'$ to the qubit $g_1$. Next, the asymmetric W state $|W_a\rangle_{ghj}$ will be taken
as the quantum channel to fulfill the sharing process. At this time, the state of the
composite system consisting of the channel qubits $(g, h, j)$ and the initial target qubit $g_1$ is
\begin{eqnarray}
|\Theta\rangle_{g_{1}ghj}&=&(\Omega|\Psi\rangle_{g_{1}})|W_a\rangle_{ghj}  \nonumber \\
&=& \frac{1}{\sqrt{2}}(a'|0\rangle_{g_1}+b'|1\rangle_{g_1})(\alpha|001\rangle_{ghj} + \beta|010\rangle_{ghj} + |100\rangle_{ghj}) \nonumber \\
&=&\frac{1}{2}\{|\psi^+\rangle_{g_1g} [a'(\alpha|01\rangle+\beta|10\rangle)_{hj}+b'|10\rangle_{hj}]+|\psi^-\rangle_{g_1g} [a'(\alpha|01\rangle+\beta|10\rangle)_{hj}-b'|10\rangle_{hj}]  \nonumber \\
&&+|\varphi^+\rangle_{g_1g} [a'|00\rangle_{hj}+b'(\alpha|01\rangle+\beta|10\rangle)_{hj}]
+|\varphi^-\rangle_{g_1g} [a'|00\rangle_{hj}-b'(\alpha|01\rangle+\beta|10\rangle)_{hj}]\}.
\end{eqnarray}

To achieve the final sharing, Grey measures her qubit pair $(g_1, g)$ in complete Bell-state bases and
notifies Holly and Jack of the measurement outcomes through classical channels (see the table 3).
Then Grey's quantum information has already been split between Holly and Jack. This can be
easily seen from equation (14). At this point, Holly and Jack perform a peculiar collective unitary
operation $U$ on qubits $h$ and $j$. The unitary operator $U$ under the ordering bases
$\{|00\rangle, |01\rangle, |10\rangle, |11\rangle\}$ takes the following form
\begin{equation}
U=
\left(
\begin{array}{cccc}
1          & 0          & 0            & 0  \\
0          & -\beta     & \alpha       & 0  \\
0          & \alpha     & \beta        & 0  \\
0          & 0          & 0            & 1
\end{array}
\right).
\end{equation}
After the collective operation, the right hand of equation (12) is transformed to
\begin{eqnarray}
\left\{ \begin{array}{*{20}{l}}
a'(\alpha|01\rangle+\beta|10\rangle)_{hj}+b'|00\rangle_{hj} \Longrightarrow a'|10\rangle_{hj}+b'|00\rangle_{hj} = \sigma^x_h(\Omega|\Psi\rangle)_h|0\rangle_j,  \\
a'(\alpha|01\rangle+\beta|10\rangle)_{hj}-b'|00\rangle_{hj} \Longrightarrow a'|10\rangle_{hj}-b'|00\rangle_{hj} = \sigma^y_h(\Omega|\Psi\rangle)_h|0\rangle_j,  \\
a'|00\rangle_{hj}+b'(\alpha|01\rangle+\beta|10\rangle)_{hj} \Longrightarrow a'|00\rangle_{hj}+b'|10\rangle_{hj} = (\Omega|\Psi\rangle)_h|0\rangle_j, \\
a'|00\rangle_{hj}-b'(\alpha|01\rangle+\beta|10\rangle)_{hj} \Longrightarrow a'|00\rangle_{hj}-b'|10\rangle_{hj} = \sigma^z_h(\Omega|\Psi\rangle)_h|0\rangle_j.
\end{array} \right.
\end{eqnarray}
Then Holly can get the conceivable state $\Omega|\Psi\rangle$
by applying one of the unitary transformations $\{\sigma_x, \sigma_y, I, \sigma_z\}$.
Hence the success probability of this scheme is $1$.

\vskip 0.4cm
\begin{center}
\begin{minipage}{400pt}
Table 3. Summary of the second stage in the second scheme.
The same as table 1 except   NUO: the necessary unitary operation.
HJS: the state on qubits $h$ and $j$ after the necessary unitary operation.
See text for more details.\\
\end{minipage}
\begin{tabular*}{13.5cm}{@{\extracolsep{\fill}}ccccccc}
\hline
GMO                         & GM  & HJCS       & NUO       & HJS  & JO          & HRQS                    \\ \hline
$|\psi^+\rangle_{g_1g}$     & 00  & $a'(\alpha|01\rangle+\beta|10\rangle)_{hj}+b'|00\rangle_{hj}$  & $U_{hj}$  & $\sigma^x_h(\Omega|\Psi\rangle)_h|0\rangle_j$
& $\sigma^x_h$  & $\Omega|\Psi\rangle_h$ \\
$|\psi^-\rangle_{g_1g}$     & 01  & $a'(\alpha|01\rangle+\beta|10\rangle)_{hj}-b'|00\rangle_{hj}$  & $U_{hj}$  & $\sigma^y_h(\Omega|\Psi\rangle)_h|0\rangle_j$
& $\sigma^y_h$  & $\Omega|\Psi\rangle_h$ \\
$|\varphi^+\rangle_{g_1g}$  & 10  & $a'|00\rangle_{hj}+b'(\alpha|01\rangle+\beta|10\rangle)_{hj}$  & $U_{hj}$  & $(\Omega|\Psi\rangle)_h|0\rangle_j$
& $I$           & $\Omega|\Psi\rangle_h$ \\
$|\varphi^-\rangle_{g_1g}$  & 11  & $a'|00\rangle_{hj}-b'(\alpha|01\rangle+\beta|10\rangle)_{hj}$  & $U_{hj}$  & $\sigma^z_h(\Omega|\Psi\rangle)_h|0\rangle_j$
& $\sigma^z_h$  & $\Omega|\Psi\rangle_h$ \\
\hline
\end{tabular*}
\end{center}

If Holly and Jack agree to recover the conceivable state $U|\Psi\rangle$ in Jack's qubit $j$,
they only need to perform a necessary collective unitary operation $U'$ instead of the original $U$
on their two qubits [cf. figure 1 (I) and (II)]. Then Holly and Jack should exchange their performances.
Concisely speaking, after the necessary collective unitary operation $U'$, Jack implements a peculiar
inverse operation to recover the conceivable state $\Omega|\Psi\rangle$, where $U'$ takes the following form
\begin{equation}
U'=
\left(
\begin{array}{cccc}
1          & 0          & 0            & 0  \\
0          & -\beta     & \alpha       & 0  \\
0          & \alpha     & \beta        & 0  \\
0          & 0          & 0            & 1
\end{array}
\right).
\end{equation}
\vskip 0.3cm

\noindent {\bf 3 \ Discussions and comparisons}

Now let us simply analyze the security of our schemes. As mentioned before, the precondition of our QOS
schemes are the quantum channels are assumed secure. Hence, it is necessary to consider the issue of
the quantum channel security before the legitimate users' actions. In our schemes the qubit distributions
through quantum channels are fulfilled by the sender Grey solely. The present quantum channels are very
similar to those in Refs.[28-34] to some extent. Consequently, the mature sampling method combining with
the block transmission technique originally presented by Long et al[8] can be used to check the quantum
channels security against any potential outside eavesdropping. As for as the potential inside cheat(s),
the strategies [28-34,35-36] including the famous decoy-photon ones proposed by C. Y. Li et al[35-36] have
been adopted in QSS schemes. They can be directly employed in our schemes and also very valid and efficient, too.
Here we do not repeat them anymore.

Let us move to reveal some features of our schemes. In both schemes, either Holly or Jack can
finally recover the conceivable state $\Omega|\Psi\rangle$. Alternatively, the concerned operation has been finally
shared by the two sharers. Hence, as for as the sharers are concerned, both schemes are symmetric.
From the description of both schemes one is readily to see that,  the first scheme
(a W state is used) is a probabilistic one, while the second scheme (an asymmetric W state is used)
is a deterministic one.

In the following let us compare our two schemes from the following five aspects: quantum
resource consumption, difficulty or intensity of necessary operations, classical resource consumption,
success probability and intrinsic efficiency. We have summarized both schemes in table 4 from these aspects.
The intrinsic efficiency of the QOS scheme is defined[37] as $\eta=P/(q+t)$,
where $q$ is the number of the qubits which are used as quantum channels (except for those chosen
for security checking), $t$ is the classical bits transmitted,  and $P$ is the final success probability.
\vskip 0.3cm
\begin{center}
\begin{minipage}{400pt}
Table 4. Comparisons between our present two schemes.  QR: quantum resource.    NO: necessary operations.
CRC: classical resource consumption.  BM: Bell state measurement.   SM: single-qubit measurement.
NCUO: necessary collective unitary operation.    SQUO: single-qubit unitary operation. \\
\end{minipage}
\vskip -0.3cm
\begin{tabular*}{13.5cm}{@{\extracolsep{\fill}}cccccc}
\hline
S           & QR              & NO                                  & CRC          &P        & $\eta$ \\ \hline
S1          & BS,$W_s$        & 2 BMs, 2 SMs, SQUO                  & 5 cbits      & 2/3     &   1/15 \\
S2          & BS,$W_a$        & 2 BMs, 2 SMs, NCUO                  & 4 cbits      & 1       &   1/9  \\
\hline
\end{tabular*}
\vskip 0.4cm
\end{center}

From table 4, one can see that both schemes consume the same quantum resources. As for the operation
complexities of the two schemes, they are almost same except that the necessary collective unitary operation
in the S2 scheme is a little difficult than the single-qubit unitary operation in the S1 scheme.
The classical resource consumption in the S2 scheme is less than that in the S1 scheme.
The success probability and efficiency of the S1 scheme are $2/3$ and $1/15$, while the S2 scheme are $1$
and $1/9$. Obviously, our second scheme is more superior and economic than the first one.

At last, we want to point out that, if the two sharers (Holly and Jack) are regarded as a single party,
then our QOS schemes are reduced to the corresponding QOT schemes. In this sense,
one can think our present schemes are more general.\\

\noindent {\bf 4 \ Summary}

To summarize, in this paper we have presented two three-party QOS schemes (S1 and S2)
for sharing a single-qubit operation on a remote qubit in either sharer's site.
Two groups of shared entanglements, i.e., a Bell and a symmetric W states,
and a Bell and an asymmetric W states, are used respectively.
Both schemes are symmetric as far as the sharers are concerned.
The S1 scheme is probabilistic while the S2 scheme is deterministic.
They consume same quantum resources. However, the S2 scheme consumes less classical resource.
The necessary operations in the S1 scheme is a little easier than those in the S2 scheme.
Hence, in all the inherent efficiency of the S1 scheme is lower than that of the S2 scheme.
Integrating all, we think the S2 scheme is better than the S1 scheme.  \\

\noindent {\bf Acknowledgements }

Supported by the Specialized Research Fund for the Doctoral Program of
Higher Education under Grant No.20103401110007, the NNSFC under Grant
Nos.10874122, 10975001, 51072002 and 51272003, the Program for Excellent
Talents at the University of Guangdong province (Guangdong Teacher Letter
[1010] No.79), and the 211 Project of Anhui University.\\

\noindent {\bf References }

\noindent[1] C. H. Bennett and G. Brassard, in Proceedings of IEEE International Conference on Computers, Systems, and Signal Processing, Bangalore, India (1984); IEEE, New York (1984) pp. 175.

\noindent[2] A. K. Ekert. Phys. Rev. Lett. 67 (1991) 661.

\noindent[3] X. H. Li, F. G. Deng and H. Y. Zhou. Phys. Rev. A 78 (2008) 022321.

\noindent[4] C. H. Bennett, et al. Phys. Rev. Lett. 70 (1993) 1895.

\noindent[5] A. Karlsson and M. Bourennane. Phys. Rev. A 58 (1998) 4394.

\noindent[6] L. Vaidman. Phys. Rev. A 49 (1993) 1473.

\noindent[7] C. Y. Cheung and Z. J. Zhang. Phys. Rev. A 80 (2009) 022327.

\noindent[8] L. Xiao, G. L. Long, F. G. Deng and J. W. Pan. Phys. Rev. A 69 (2004) 052307.

\noindent[9] T. J. Wang and F. G. Deng. Commun. Theor. Phys. 55 (2011)795.

\noindent[10] M. Hillery, V. Buzek and A. Berthiaume. Phys. Rev. A 59 (1999) 1829.

\noindent[11] A. Karlsson, M. Koashi and N. Imoto. Phys. Rev. A 59 (1999) 162.

\noindent[12] H. Yuan, Y. M. Liu, L. F. Han and Z. J. Zhang. Commun. Theor. Phys. 49 (2008) 1191.

\noindent[13] S. F. Huelga, J. A. Vaccaro and A. Chefles. Phys. Rev. A 63 (2001) 042303.

\noindent[14] A. M. Wang. Phys. Rev. A 74 (2006) 032317.

\noindent[15] A. M. Wang. Phys. Rev. A 75 (2007) 062323.

\noindent[16] Y. S. Zhang, M. Y. Ye and G. C. Guo. Phys. Rev. A 71 (2005) 062331.

\noindent[17] Z. J. Zhang and C. Y. Cheung. J. Phys. B 44 (2011) 165508.

\noindent[18] S. F. Wang, Y. M. Liu, J. L. Chen, X. S. Liu and Z. J. Zhang. Quantum Inf. Process. (DOI 10.1007/s11128-013-0537-5)(2013).

\noindent[19] B. L. Ye, Y. M. Liu, X. S. Liu and Z. J. Zhang. Chin. Phys. Lett. 2 (2013) 020301.

\noindent[20] D. C. Liu, Y. M. Liu, X. F. Yin, X. S. Liu and Z. J. Zhang. Quantum inf. Process. (Under review)

\noindent[21] Q. B. Ji, Y. M. Liu, X. F. Yin, X. S. Liu and Z. J. Zhang. Quantum Inf. Process. (DOI 10.1007/s11128-013-0533-9) (2013).

\noindent[22] D. C. Liu, Y. M. Liu, X. F. Yin, X. S. Liu and Z. J. Zhang. Int. J. Quant. Information (accepted) (2013).

\noindent[23] W. Dur, G. Vidal and J. I. Cirac. Phys. Rev. A 62 (2000) 062314.

\noindent[24] S. B. Zheng. Phys Rev A 74 (2006) 054303.

\noindent[25] X. Q. Zuo, Y. M. Liu, W. Zhang, X. F. Yin and Z. J. Zhang. Int. J. Quantum Inf. 6 (2008) 1245.

\noindent[26] G. X. Pan, et al. Common. Theor. Phys. 51 (2009) 227.

\noindent[27] Y. M. Liu, X. F. Yin, W. Zhang and Z. J. Zhang. Int. J. Quantum Inf. 7 (2009) 349.

\noindent[28] G. L. Long, X. S. Liu, Phys. Rev. A 65 (2002) 032302.

\noindent[29] Z. J. Zhang, J. Yang, Z. X. Man and Y. Li. Eur. Phys. J. D 33 (2005) 133.

\noindent[30] Z. J. Zhang, Y. Li and Z. X. Man. Phys. Rev. A 71 (2005) 044301.

\noindent[31] Z. J. Zhang and Z. X. Man. Phys. Rev. A 72 (2005) 022303.

\noindent[32] F. G. Deng, X. H. Li, C. Y. Li, P. Zhou and H. Y. Zhou. Phys. Rev. A 72 (2005) 044301.

\noindent[33] F. G. Deng, X. H. Li, H. Y. Zhou and Z. J. Zhang. Phys. Rev. A 72 (2005) 044302.

\noindent[34] Z. J. Zhang. Phys. Lett. A 342 (2005) 60.

\noindent[35] C. Y. Li, H.Y. Zhou, Y. Wang and F. G. Deng. Chin. Phys. Lett. 22 (2005) 1049.

\noindent[36] C. Y. Li, X. H. Li, F. G. Deng, P. Zhou, Y. J. Liang and H. Y. Zhou. Chin. Phys. Lett. 23 (2006) 2896.

\noindent[37] H. Yuan, Y. M. Liu, W Zhang and Z. J. Zhang. J. Phys. B 41 (2008) 145506.
\end{document}